\documentclass{elsart}
\usepackage{amstext}
\usepackage{graphicx}
\usepackage{epsfig}
\begin{document}
\begin{frontmatter}
\title{np-nh bands in the N=28 isotones}
\author{A. Poves and J. S\'anchez Solano}
\address{Departamento de F\'{\i}sica Te\'orica C-XI, Universidad
  Aut\'onoma de Madrid, E--28049 Madrid, Spain}
\author{E. Caurier and F. Nowacki}
\address{Groupe de Physique Th\'eorique, IReS, B\^at. 27, 
    IN2P3-CNRS/Universit\'e Louis Pasteur, BP 28,
  F--67037 Strasbourg Cedex~2, France}

\begin{abstract}
The
existence of n-particle 
n-hole deformed yrare bands in the N=28
isotones is explored using full $pf$-shell diagonalizations and the
Lanczos Strength Function method. We find different 2p-2h and
4p-4h collective bands that, when allowed to mix, more often
disappear. Only the 2p-2h yrare band in $^{52}$Cr and the 4p-4h yrare
band in $^{56}$Ni survive, and only  in this latter case, due
to the reduced density of 2p-2h states, can the band be seen as a
$\gamma$-cascade.
\end{abstract}
\end{frontmatter}


The occurrence of yrare bands of enhanced collectivity, with a well
defined particle-hole structure on top of the ground state, has been
recently documented by experiments in $^{56}$Ni \cite{rudolf},
$^{36}$Ar \cite{carlsv} and $^{40}$Ca \cite{idegu}. In the Ni and Ar 
cases the bands are dominantly made
of 4p-4h excitations while in the $^{40}$Ca the leading structure is
8p-8h \cite{ca40sd}. In $^{56}$Ni the band is clasified as highly
deformed $\beta$= 0.3/0.4 , while in $^{36}$Ar and $^{40}$Ca they are
 characterised as superdeformed with  $\beta$= 0.4/0.5 and  $\beta$=
 0.5/0.6 respectively. In
a recent study \cite{mizu}, Mizusaki and co-workers have carried out a
theoretical search for this kind of bands in the N=28 isotopes, 
using  the Monte Carlo shell model approach. In this article we 
examine the same topic, in the framework of full $pf$-shell
diagonalizations using the Lanczos method.

The $pf$-shell has been the locus of a lot of activity in nuclear
structure since the advent of the exact 0$\hbar \omega$
calculations in $^{48}$Cr~\cite{a48}.
 In a recent  article \cite{A50} we have extend the full $pf$-shell
 calculations up to A=52 and we have introduced the interaction KB3G,
 the one used in this work. 
 The diagonalizations are performed in the $m$-scheme using a fast
 implementation of the Lanczos algorithm through the code {\sc
 antoine}~\cite{antoine} or in  J-coupled scheme using the code
 {\sc nathan} \cite{nathan}. Some details may be found in
 ref.~\cite{masses}.  The Lanczos Strength Function (LSF) method,
 following  Whitehead's prescription~\cite{white}, is explained and 
 illustrated in refs.~\cite{cpz1,cpz2,bloom}. 


 The calculation of yrare bands in nuclei with  large shell model
 dimensionalities, poses serious computational problems. The reason
 is that, most often, the interesting states lie at  excitation
 energies  where the level density is high. This represents a real
 challenge for all the shell model approaches, because it requires the
 calculation of many states of the same spin and parity. While this
 can be easily done for m-scheme dimensions of  a few millions, the
 task becomes formidable for dimensions of tens of millions. The
 method of  ref~\cite{mizu} relies in the existence of local minima in
 the projected energy surface resulting of a constrained Hartree-Fock
 calculation. The Slater determinants corresponding to these minima
 are then fed into the quantum Monte Carlo diagonalization method
 ($qmcd$).
  While
 this method is reasonably under control for the yrast states and
 other states originating in  well defined minima, it is not clear how
 would it work in less clear-cut situations. As an example,
 ref~\cite{mizu} claims that the $qmcd$ method is unpractical to describe
 a possible  4p-4h band in $^{54}$Fe, which is then studied at a lower
 level of approximation, doing variation after projection (VAP) on the
 cranked Hartree-Fock solution. 

 Our method proceeds by three steps:
 
 I) The space of solutions is explored 
 by means of diagonalizations  restricted to configurations with
 a fixed number of particle-hole excitations on the top of the
 reference one (1f$_{7/2}$)$^N$, 
 searching for collective bands.

 II) If such bands  are found, we proceed to
 immerse the  bandhead state in the full space by
 means of the Lanczos Strength Function method, i.e. we take the np-nh
 ground state and use it as starting vector for the Lanczos iterations
 in the full space, in order to know whether it
 stands the mixing or not.

 III) If it survives to the mixing, we try 
 to build the physical band acting repeatedly upon it  with the
 quadrupole operator and using  again the LSF method as described below.

 Let's proceed with the first step in the four  cases of interest,
 $^{50}$Ti, $^{52}$Cr, $^{54}$Fe and  $^{56}$Ni. In
 tables~\ref{tab:ti50ph},~\ref{tab:cr52ph},~\ref{tab:fe54ph}
 and~\ref{tab:ni56ph} we list the excitation energies and the BE2's
 corresponding to the fixed 2p-2h and 4p-4h bands. 
 In $^{50}$Ti none of the bands appear  to be really collective, even
 though there is a certain quadrupole linking among the states. 
 In  $^{52}$Cr both are quite regular, with large BE2's in the 4p-4h
 case. For $^{54}$Fe  the situation is more complex; the 2p-2h band is
 clearly less collective than the corresponding one in  $^{52}$Cr,
 with a neat bifurcation at J=6,
 while the 4p-4h band mimics perfectly a well deformed rotor. The same
 is valid for the 4p-4h band of  $^{56}$Ni; in this nucleus the 2p-2h band
 shows no collectivity at all. 
\begin{table}[h]
\begin{center}
\caption{Properties of the fixed 2p\,-2h and 4p\,-4h bands in
 $^{50}$Ti. Energies in MeV,  B(E2)'s in $e^2\,fm^4$,
 Q's in $e\,fm^2$.}
\label{tab:ti50ph}
\vspace{0.3cm}
\begin{tabular*}{\textwidth}{@{\extracolsep{\fill}}ccccc}
\hline\hline
     &\multicolumn{2}{c}{2p\,-2h}  &\multicolumn{2}{c}{4p\,-4h} \\
\cline{2-3}\cline{4-5}
J & $\Delta$E & B(E2)$\downarrow$ & $\Delta$E  & B(E2)$\downarrow$ \\
\hline
0 & 0.00 &     & 0.00 & \\
2 & 0.54 & 78  & 0.59 & 126 \\
4 & 1.28 & 106 & 1.79 & 153 \\
6 & 1.99 & 103 & 2.42 & 1 \\
8 & 2.87 & 86  & 3.49 & 122 \\
10& 4.15 & 66  & 4.67 & 80 \\
12& 5.59 & 40  & 5.97 & 51 \\
\hline\hline
\end{tabular*}
\end{center}
\end{table}

\begin{table}[h]
\begin{center}
\caption{Properties of the fixed 2p\,-2h and 4p\,-4h bands in
 $^{52}$Cr. Energies in MeV,  B(E2)'s in $e^2\,fm^4$,
 Q's in $e\,fm^2$.}
\label{tab:cr52ph}
\vspace{0.3cm}
\begin{tabular*}{\textwidth}{@{\extracolsep{\fill}}ccccc}
\hline\hline
     &\multicolumn{2}{c}{2p\,-2h}  &\multicolumn{2}{c}{4p\,-4h} \\
\cline{2-3}\cline{4-5}
J & $\Delta$E & B(E2)$\downarrow$ & $\Delta$E  & B(E2)$\downarrow$ \\
\hline
0 & 0.00 &     & 0.00 & \\
2 & 0.39 & 112  & 0.46 & 242 \\
4 & 1.08 & 157 & 1.21 & 313 \\
6 & 2.00 & 162 & 1.97 & 322 \\
8 & 3.12 & 156  & 2.92 & 325 \\
10& 4.31 & 105  & 4.33 & 304 \\
12& 5.60 & 96  & 6.10 & 241 \\
\hline\hline
\end{tabular*}
\end{center}
\end{table}

\begin{table}[h]
\begin{center}
\caption{Properties of the fixed 2p\,-2h and 4p\,-4h bands in
 $^{54}$Fe. The numbers in
 parenthesis correspond to non-yrast states that appear to be more
 strongly related to the band. Energies in MeV,  B(E2)'s in $e^2\,fm^4$,
 Q's in $e\,fm^2$.}
\label{tab:fe54ph}
\vspace{0.3cm}
\begin{tabular*}{\textwidth}{@{\extracolsep{\fill}}ccccc}
\hline\hline
     &\multicolumn{2}{c}{2p\,-2h}  &\multicolumn{2}{c}{4p\,-4h} \\
\cline{2-3}\cline{4-5}
J & $\Delta$E & B(E2)$\downarrow$ & $\Delta$E  & B(E2)$\downarrow$ \\
\hline
0 & 0.00 &     & 0.00 & \\
2 & 0.45 & 101  & 0.35 & 320 \\
4 & 1.35 & 138 & 1.07 & 447 \\
6 & 2.36(2.51) & 11(125) & 2.09 & 479 \\
8 & 3.61(3.82) & 44(91)  & 3.36 & 478 \\
10& 4.79 & 1  & 4.86 & 466 \\
12& 4.92 & 11 & 6.56 & 416 \\
\hline\hline
\end{tabular*}
\end{center}
\end{table}

\begin{table}[h]
\begin{center}
\caption{Properties of the fixed 2p\,-2h and 4p\,-4h bands in
 $^{56}$Ni. Energies in
 MeV,  B(E2)'s in $e^2\,fm^4$, Q's in $e\,fm^2$.}
\label{tab:ni56ph}
\vspace{0.3cm}
\begin{tabular*}{\textwidth}{@{\extracolsep{\fill}}ccccc}
\hline\hline
     &\multicolumn{2}{c}{2p\,-2h}  &\multicolumn{2}{c}{4p\,-4h} \\
\cline{2-3}\cline{4-5}
J & $\Delta$E & B(E2)$\downarrow$ & $\Delta$E  & B(E2)$\downarrow$ \\
\hline
0 & 0.00 &    & 0.00 & \\
2 & 0.74 & 25 & 0.43 & 313 \\
4 & 1.50 & 13 & 1.37 & 439 \\
6 & 1.79 & 24 & 2.70 & 467 \\
8 & 1.82 & 1  & 4.30 & 459 \\
10& 2.74 & 3  & 6.07 & 418 \\
12& 4.99 & 0  & 7.90 & 283 \\
\hline\hline
\end{tabular*}
\end{center}
\end{table}

\begin{figure}[h]
\begin{center}
    \epsfig{file=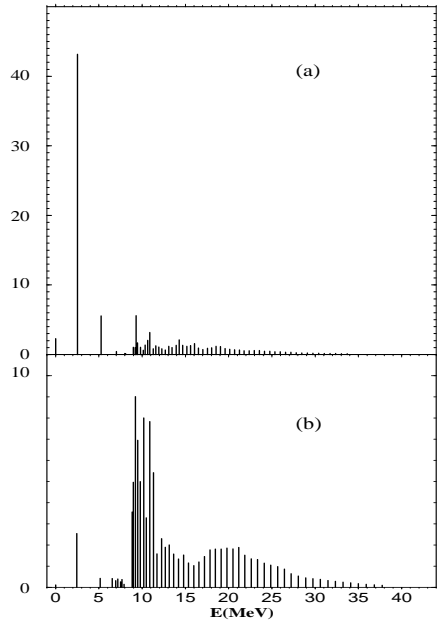,height=9cm,width=7cm}
    \caption{Overlaps of the 2p-2h(a) and 4p-4h(b) band-heads in
    $^{52}$Cr with the physical states.}
    \label{fig:fig35}
\end{center}
\end{figure}

 In our second step, we take the 0$^+$ band-head and allow it to
 mix with all the states. To optimize the calculation as well as to
 keep track of the initial state, we take it as starting vector
 (``pivot'') in the Lanczos construction in the full space (or in a
 large enough one). After N iterations, we compute its  overlaps with
 the N quasi-physical states. For the 
 fixed np-nh bandhead to be representative of a ``physical''
 bandhead, there must be a single state that has a large overlap with
 it. As a  bonus, this state is  converged in very
 few iterations, what makes the method computationally efficient.
 In figure~\ref{fig:fig35} we have plotted the overlaps for the
 2p-2h (upper panel) and 4p-4h (lower panel) band-heads in $^{52}$Cr. The 
 calculations are carried out in the full $pf$-shell (m-scheme dimension 45
 millions) using the coupled code Nathan (0$^+$ dimension, 8 $\times$
 10$^5$) and 150 iterations are made. Notice how the 2p-2h state
 is concentrated at 45\% in one ``physical''  state with no other one
 taking any relevant share. The excitation energy of this state,
 2.43~MeV, is in correspondence with the experimental 0$^+_2$ at
 2.65~MeV. It has a 46\% content of 2p-2h, 26\% of 3p-3h,
  18\% of 4p-4h and minor percentages of the other components. The
 fact that the mixing takes place with components of higher p-h rank
 should result in an increase of collectivity with respect to the
 fixed 2p-2h band. Besides the ``coherent'' 2p-2h state, the
 next four 0$^+$ states produced by the calculation are also in
 correspondence with the  experimental 0$^+$'s:
\begin{center} 
 0$^+_3$; exp. 4.74~MeV; th. 5.15~MeV,

 0$^+_4$; exp. 5.60~MeV; th. 5.47~MeV,

 0$^+_5$; exp. 5.75~MeV; th. 6.04~MeV,

 0$^+_6$; exp. 6.10~MeV; th. 6.30~MeV,
\end{center}
 a rather spectacular achievement of the shell model spectroscopy!!

 On the contrary, the much more collective
 4p-4h band-head is completely dissolved in the full space; no
 physical state carries a dominant fraction of the overlap. According
 to that, no 4p-4h band exists in $^{52}$Cr. Whether the 2p-2h band has any
 experimental relevance or not, will be discussed below. 

  Very much the same occurs in $^{50}$Ti; the 2p-2h bandhead survives
  at 61\% in the full space 0$^+_2$ at 4.08~MeV (exp. at 3.87~MeV);
  the 4p-4h bandhead is fully dispersed among many physical states.

\begin{figure}[h]
\begin{center}
    \epsfig{file=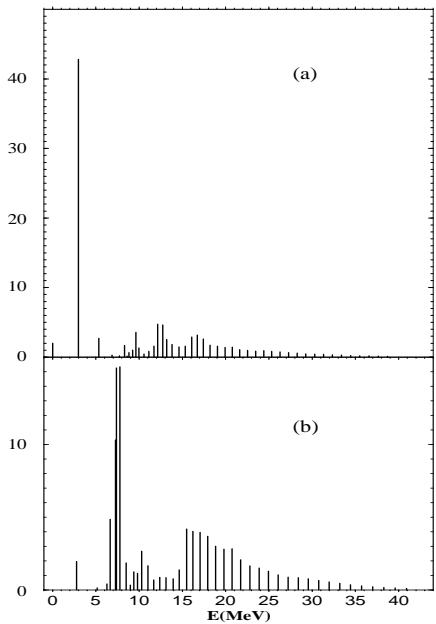,height=9cm,width=7cm}
    \caption{Overlaps of the 2p-2h(a) and 4p-4h(b) band-heads in
    $^{54}$Fe with the physical states.}
    \label{fig:fig36}
\end{center}
\end{figure}

  For $^{54}$Fe the full space calculations of the overlap
  distribution become unaffordable. However, we think that it is
  sufficiently safe to compute the np-nh mixing at a level of truncation
  t=n+4. In figure~\ref{fig:fig36} we present
  the results of these calculations (t=6 for the 2p-2h -upper panel-
  and t=8 for the 4p-4h -lower panel-). In this last case, the
  m-scheme dimension is 177
  millions. Using the coupled code Nathan we deal with a 0$^+$
  dimension of 2.5 $\times$ 10$^6$ and make 150 iterations. The
  situation resembles that of $^{52}$Cr; the coherent 2p-2h
  state represents 43\% of
  the  0$^+_2$ at 2.77~MeV (exp. at 2.56~MeV) and no other state
  carries a significant fraction of it. In the 4p-4h case the
  fragmentation is complete, no residue of a band is left. According
  to this result, the claim in ref.~\cite{mizu} of the existence of a
  4p-4h band in this nucleus, based on a VAP-CHF calculation, has to be
  taken with  some skepticism.

  For $^{56}$Ni we refer to our calculations in \cite{rudolf} and 
  to the experimental results to state the existence of a
  4p-4h band. The reason why it survives in this nucleus and not in
  $^{54}$Fe, is clearly the  very different level density in both
  nuclei. In  $^{56}$Ni, because it is doubly magic, the number of
  2p-2h states is nearly an order
  of magnitude smaller that in $^{54}$Fe, whose 4p-4h bands wrecks in
  a  sea of 2p-2h states.

   This brings us to step III.
  We have now several candidates to bandheads, but, do they develop a real
  band? The answer is ``yes'' in the well known case of
  the 4p-4h band in $^{56}$Ni. We shall now proceed to examine  the
  situation in the other cases. 
  It could be done ``brute force'' by computing many
  states of each spin and their E2 transition probabilities
  and looking for band-like patterns, but this approach is only
  realistic (from the computational point of view) in
  $^{50}$Ti, where we have actually used it to check our approximations.
  The method we use is, once again, an application of the Lanczos
  strength function technique. We start with the bandhead 0$^+$ and
  apply the quadrupole operator on it: $Q | 0^+ \rangle$. This
  operation generates a
  threshold vector that we take as the ``pivot'' in the Lanczos
  procedure. We perform N$\sim$100 iterations to produce 
  an approximate strength
  function. At this stage, either there is one state that carries most
  of the strength, signaling its pertenence to the band, or not, in
  which case there is no band. In the positive case the resulting
   $|2^+\rangle$
  state is retained as a band member and we proceed to act  with $Q$
  on it. Now there are  several
  possible angular momentum couplings,
  but we follow the  $\Delta$J=2 path in the
  even-even nuclei. The procedure is repeated until the strength
  bifurcates, dilutes or plainly disappears.

  In $^{50}$Ti we obtain a sequence of states that fulfill the
  requirements discussed above. However, their energy spacings are not
  very much rotor-like and their B(E2)'s are only about 100 $e^2\, fm^4$. 
  These values are comparable to those of the yrast band. Some of these states
  have been measured, but they
  do not appear to show any band signature.

\begin{table}[b]
\begin{center}
\caption{Yrare 2p\,-2h  band in $^{52}$Cr. Excitation energies are
  relative to the 0$^+_2$ band-head, experimentally at 2.64~MeV and
  calculated at 2.43~MeV. Energies in MeV,  B(E2)'s in $e^2\,fm^4$,
  Q's in $e\,fm^2$}
\label{tab:cr52yrare}
\vspace{0.3cm}
\begin{tabular*}{\textwidth}{@{\extracolsep{\fill}}cccccc}
\hline\hline
J & $\Delta$E & \% of B(E2)$\uparrow$ & \% of 2p\,-2h &
B(E2)$\downarrow$ & Q$_{spec}$ \\
\hline
0 & 0.00 &    & 46 &     &      \\
2 & 0.44 & 77 & 43 & 238 & -27.7\\
4 & 1.32 & 82 & 43 & 323 & -39.3\\
6 & 2.51 & 77 & 42 & 329 & -39.3\\
8 & 4.14 & 83 & 44 & 331 & -39.2\\
10& 6.03 & 66 & 39 & 239 & -25.5\\
12& 8.02 & 27 & 51 &  66 & -13.7\\
12& 8.31 & 28 &    &  69 &      \\
\hline\hline
\end{tabular*}
\end{center}
\end{table}

  In  $^{52}$Cr we can definitely speak of a ``theoretical'' yrare
  band. Our results for the excitation energies are very close to
  those of ref.~\cite{mizu}. In
  table \ref{tab:cr52yrare} we list  the
  excitation energies, the percentage of the B(E2)$\uparrow$ of the
  transition from J-2 to J carried by each state, the amount of 2p-2h
  components, the B(E2)$\downarrow$ and the spectroscopic quadrupole
  moments. All these numbers are consistent with a deformed band up to
  J=10, afterwards a bifurcation and reduction of the strength takes
  place. Notice the large B(E2) values, similar to those of the yrast
  band of $^{48}$Cr, corresponding to $\beta \sim$~0.3. Besides, the
  energy spacings are very close to those of a rigid rotor;
 
\medskip
\begin{center}
 $\displaystyle{\frac{\Delta E(4^+)}{\Delta E(2^+)}}$=3(3.33);

 $\displaystyle{\frac{\Delta E(6^+)}{\Delta E(4^+)}}$=1.9(2.1);  

 $\displaystyle{\frac{\Delta E(8^+)}{\Delta E(6^+)}}$=1.65(1.71);  

 $\displaystyle{\frac{\Delta E(10^+)}{\Delta E(8^+)}}$=1.46(1.53);  
\end{center}
\medskip

\noindent
 (the values in parenthesis are those of the rigid rotor limit). The
 percentage of the 2p-2h components as well as the occupation
 numbers of the individual orbits are nicely constant too. The
 spectroscopic quadrupole moments correspond to 
 a prolate rotor and are fully consistent with the B(E2)'s.
 The sequence of calculated states, 0$^+$ at 2.43~MeV, 2$^+$ at
 2.87~MeV, 4$^+$ at 3.75~MeV and 6$^+$ at 4.94~MeV can be put in
 correspondence with the experimental ones at  2.64~MeV (0$^+$), 
 3.16~MeV (2$^+$),  4.04~MeV (4$^+$) and  5.14~MeV (6$^+$). However,
 there are no transitions experimentally observed linking these
 states. This means that, due to the presence of many other 2p-2h
 states to which to decay by M1 transitions and to the phase space
 enhancement of the E2 transitions to the yrast band, the
 ``theoretical'' yrare band does not show up experimentally as
 a $\gamma$-cascade. 
 In $^{54}$Fe the situation is even worse. The bifurcation, already
 present at the fixed 2p-2h level, appears also in the ``physical''
 band. Only the sequence 0$^+$, 2$^+$, 4$^+$ behaves like a band, but
 with B(E2)'s substantially smaller than in the $^{52}$Cr yrare band.

 In summary, we have found that yrare bands of a definite np-nh
 character exist in the N=28 isotones. When full mixing is permitted,
 only the 2p-2h bands survive in   $^{50}$Ti,  $^{52}$Cr and  $^{54}$Fe,
 with different degrees  of collectivity. In  $^{56}$Ni it is the 4p-4h
 band which is physical. This is due to the difference in the 2p-2h level
 density in the doubly magic case. In addition, we find that the 2p-2h
 bands preferentially decay out, making their experimental
 identification very difficult.

\ack
 This work has been partly supported by MCyT~(Spain), grant BFM2000-30
 and by the IN2P3~(France)-CICyT~(Spain)
 agreements. We also thank the CCC-UAM for a computational grant.

\end{document}